\documentstyle[prl,aps]{revtex}
\def\alumina{$\alpha$-Al$_2$O$_3$}
\begin{document}
\title{Molecular Dynamics Simulation of Electron Trapping in
the Sapphire Lattice}
\author{
C. Rambaut~$^*$, K. H. Oh~$^*$, H .Jaffrezic~$^\#$,
J. Kohanoff~$^\&$, and S. Fayeulle~$^*$
}
\address{
{}~$^*$ Ecole Centrale de Lyon, D\'epartement Mat\'eriaux et
M\'ecanique-Physique, URA CNRS 447, 69131 Ecully Cedex, France.
}
\address{
$^\#$ Institut de Physique Nucl\'eaire IN2P3, 69341 Villeurbanne
Cedex, France.
}
\address{
$^\&$ International Centre for Theoretical Physics, 34014 Trieste, Italy.
}
\date{\today}
\maketitle
\begin{abstract}
Energy storage and release in dielectric materials can be described on
the basis of the charge trapping mechanism. Most phenomenological aspects
have been recently rationalized in terms of the space charge
model~\cite{blaise,blaise1}. Dynamical aspects are studied here by
performing Molecular Dynamics simulations.
We show that an excess electron introduced into the sapphire lattice
(\alumina) can be trapped only at a limited number of sites. The energy
gained by allowing the electron to localize in these sites is of the order
of 4-5 eV, in good agreement with the results of the space charge model.
Displacements of the neighboring ions due to the implanted charge are
shown to be localized in a small region of about 5~\AA.
Detrapping is observed at 250 $K$. The ionic displacements turn out to play
an important role in modifying the potential landscape by lowering, in a
dynamical way, the barriers that cause localization at low temperature.
\end{abstract}

\twocolumn

\section*{Introduction}

Recent theoretical and experimental efforts have focused on the mechanism
of electron trapping in dielectric materials. The space charge
model~\cite{blaise,blaise1} based on the concept of polaron~\cite{mott}
has been developed to explain trapping and its consequences from an
energetic point of view. The main underlying idea is that electrons trapped
into the lattice build up a space charge simultaneously generating a
mechanical deformation (polaron). Energy is stored both in electrostatic
(coupling) and in mechanical (internal) forms~\cite{bigarre}. Beyond a
critical concentration of trapped electrons, dielectric breakdown or
flashover occurs. The mechanical part of the energy is then released
in the form of a thermal shock wave that produces a local heating large
enough to create a plasma, eventually leading to the rupture of the
material. One of the main consequences is that dielectric breakdown would
not arise because of an external driving force (e.g. a high voltage
difference between the electrodes), but because of the internal stress
caused by the accumulation of the trapped charges. Thorough investigations
have shown that electrons can be trapped in a lattice as soon as a local
variation of the dielectric susceptibility occurs. This variation can be
either due to defects (vacancies, impurities, interfaces, etc...) or to
crystallographic dissymetry. Once the charge is trapped the lattice is
distorted and polarized. The energy involved in this process has been
evaluated within some realistic approximation, and it has been shown that an
energy of at least 5 eV is stored upon electron trapping.

On the other hand, trapping has been studied experimentally using the
mirror method~\cite{valla1,valla2}. It has been shown that trapping can
occur in pure
sapphire (\alumina~single-crystals) only for temperatures lower than 250 $K$,
but the trapping/detrapping dynamics in this kind of materials is not
yet fully understood. In particular, the dynamic behavior in the range
of pico-second to femto-second cannot be observed in real experiments,
while it is crucial to elucidate the charge trapping mechanisms. In this
work we apply Molecular Dynamics technique (MD), which has widely proven
efficiency in reproducing the properties of real materials~\cite{nga}, to
explore these aspects of electron trapping in \alumina. Ionic polarization
has been described in terms of a shell model which is briefly outlined
in the following section. We then show how the solvated electron can
stay firmly in some specific lattice sites at low temperature producing
a local polarization and how, despite the fact that the potential
turns out to be quite deep, the electron can self-detrap at moderate
temperatures.

\section*{Shell Model and details of the simulation}

The potential used for corundum was based on the assumption of a fully
ionic model~\cite{catlow}. Ionic polarization was introduced through a
modified shell model. In the shell model introduced by Dick and
Overhauser~\cite{dick}, each ion is represented by a massive core and a
massless shell which simulates the valence electrons. The total ionic
charge divides amongst core and shell, the core-shell interactions being
described by harmonic springs. In this work we use a modified shell model
due to Catlow and Stoneham~\cite{catlow}.

The assumption of massless shells is a way to introduce the adiabatic
(Born-Oppenheimer) approximation in the context of the shell model. It means
that shells have to adjust instantaneously to the present configuration
of the cores. This implies that a full relaxation of the shells is
needed at each step of the MD simulation. In practice, full relaxation is
very difficult and costly to achieve, but if a less strict convergence is
required, a systematic error is accumulated during the simulation that
unphysically damps the motion of the cores~\cite{remler}. The slight
inconsistency between force and energy translates into a continuous and
systematic energy loss. To avoid this problem we adopted a
Car-Parrinello-like strategy by assigning a small, but finite mass to
the shells, and by treating
them as dynamical variables evolving according to their own (fictitious)
equations of motion~\cite{cp}. This approach proved to be very efficent in the
context of ab initio MD, and the reason is that a second order dynamics
for the shells gives rise to oscillating fictitious forces on the cores
that average out during the slow dynamics of the cores. We have chosen a mass
of 10\% the proton mass, so ensuring energy conservation to high accuracy.
The Coulombic contribution to the interionic potential is calculated
by the Ewald summation technique~\cite{hansen,allen}. Non-Coulombic,
short-range interactions between shells are described by the Buckingham
potential~\cite{bucky}, i.e.

\begin{equation}
V_{\rm ij}=A_{\rm ij}~\exp\left(-\frac{r_{\rm ij}}{\rho}\right)-
\frac{C_{\rm ij}}{r_{\rm ij}}
\end{equation}

The sets of parameters A$_{\rm ij}$, C$_{\rm ij}$, and $\rho$
for each pair of ionic species are
displayed in Table I, and the values of the spring constant and the
partial charges for cores and shells in Table II. All values have been
taken from Ref.~\cite{catly}, except for the Al$^{3+}$ core-shell spring
constant which was adjusted to fit the dielectric constant
of \alumina~computed from the simulation to the experimental value
($\epsilon_r=9.8$).

The simulation has been performed on a system of 120 ions at constant
number, volume and temperature (canonical ensemble). We have placed the
particles into an orthorhombic simulation box of experimental lattice
constants, and replicated using periodic boundary conditions. Verlet
algorithm was used to integrate the equations of motion. A time step of
0.1 fs had to be used in order to integrate properly the equations
of motion for the shells, due to their small mass. Simulations were
started with all ions at their equilibrium positions, and velocities
taken at random from a Maxwell-Boltzmann distribution at the desired
temperature. Before simulating the charge trapping we have verified the
accuracy of the model on reproducing the physical properties of pure
\alumina, namely the lattice energy and the mean square displacement. We
have found a value of -160 eV at $T=300~K$, compared to -160.4 eV from
experiment~\cite{samsonov} and -160.24 eV from Catlow and al.~\cite{catly}.
The mean square displacement also compared reasonably with experimental
measurements at $300~K$ and $2170~K$~\cite{esnouf}, as shown in Figs. 1 and
2. These results obtained with our modified shell model gave us confidence
to continue the study of charge trapping by introducing an excess electron
into the corundum \alumina~lattice.

\section*{Trapping simulation}

The lattice of alumina can be described as a succession of anionic and
cationic planes with hexagonal in-plane ordering. The O$^{2-}$ ions form
an hexagonal compact stacking, slightly distorted in order to make room for
the larger Al$^{3+}$ ions. All sites of the crystallographic arrangement
are non-equivalent~\cite{geshwind}. We first run the simulation for a
perfect crystal of 120 ions for 10000 time steps (equal to 1 ps) to ensure
thermal equilibration. To obtain a hint on possible trapping sites we have
computed the (unrelaxed) potential energy surface at the (001) plane that
contains the O$^{2-}$ ions.

\subsection*{Potential energy surface}

The potential of the uncharged lattice at a generic point (x,y) in the (001)
plane is given by the Coulomb interaction between the ions in the crystal and
a test particle of charge unity placed at that point. Since the test particle
is a hypothetical object, the potential has to be computed with all ions in
the equilibrium positions of the uncharged lattice. The resulting surface is
shown in Fig. 3. It can be osverved the existence of local minima where the
electron can be trapped. We have chosen the position of one of these potential
wells as the initial position for the excess electron in our simulation.

\subsection*{Model for the ion-electron interaction}

In this first approach to the problem of electron dynamics in dielectrics we
have considered the electron as a purely classical particle. The interaction
potential between the electron and O$^{2-}$ was modelled by a pure Coulomb
repulsion at all distances.

\begin{equation}
\Phi_{eO^{2-}}(r)=\frac{Z_j e^2}{r}\qquad{\rm for~core~and~shell}
\end{equation}

The e$^-$--Al$^{3+}$ interaction was instead described in terms of
short-range repulsion, as though the excess electron was an oxygen
shell, but of charge unity and finite mass (the electron mass).

\section*{Results}

We have performed simulations at several different temperatures up to 300 $K$.
Fig. 4 shows the path of the (classical) electron at 300 $K$, and Fig. 5
corresponds to 200 $K$. Also the potential energy surface in the (001) plane
is shown in the plots. In Fig. 6 we plot the mean square distance traveled
by the electron as a function of time. It is clearly apparent that at 300 $K$
the electron escapes from the initial trapping site, while at 200 $K$ it stays
close to the bottom of the potential well. The transition from one regime to
the other has been located around 250 $K$. Below this temperature the excess
electron solvates, and above it jumps from one well to another. These results
are in good agreement with experimental data obtained by the mirror
method~\cite{bigarre} .

In no case the kinetic energy of the ions is sufficient to overcome the
barriers present, but at high temperatures the ions in the polarization
cloud can respond more rapidly by lowering the barriers in a dynamical way.
Indeed, the active presence of the excess electron has an important
backreaction effect onto the neighboring ions. When the excess electron is
trapped, it attracts the positive Al$^{3+}$ ions and repels the negative
O$^{2-}$ ions causing a lattice distortion and a dielectric polarization.

Despite the limited size of our system we have been able to determine the
main features and magnitude of the ionic dispacements. The six nearest oxygen
ions moved significantly outwards, away from their equilibrium positions,
typically between 0.17 and 0.27~\AA. The displacement pattern of the Al$^{3+}$
ions is more complicated. The two first neighbors are very stable, with an
average vibration of 0.025~\AA. This is because these two ions are strongly
attracted towards the tapped electron, and hence they behave as though their
mass were renormalized by the interaction. The next two Al$^{3+}$ neighbors
are also largely attracted towards the trap, because the outwards motion of
the O$^{2-}$ ions leaves enough place for them to move in. The average
fluctuation of the Al$^{3+}$ ions is about $(0.07\pm 0.01)$~\AA. The outer
Al$^{3+}$ ions do not experience any significant force, as a consequence of
the short range of the e$^-$--Al$^{3+}$ interaction. According to this
pattern of distortions, we estimate the size of the polaron in about 4.8~\AA.

We have computed the potential energy surface for the charged lattice at
$T=200~K$, and we show it in Fig. 7 compared to the uncharged potential.
The former was calculated with the ions in their displaced positions after
the introduction of the excess electron, but only the potential felt by the
test charge was computed, in order to compare with the uncharged lattice.
The bottom of the potential well turns out to be 4.7 eV lower in energy in
the case of the deformed (charged) lattice. Therefore, the effective presence
of the electron further stabilizes the trap by increasing the depth of the
potential well. These results are consistent with experiments showing that
the energy necessary to detrap charging electrons is larger than the energy
gained upon trapping them~\cite{valla1}. The total energy of the lattice
increases
about 4-5 eV upon electron implantation, so that this number represents the
energy of formation of the polaron in \alumina. This value is consistent with
the estimation done using the space charge model~\cite{blaise}. Part of the
energy is stored as mechanical energy and, when dielectric breakdown occurs,
it is supposed to be released in the form of a shock thermal wave that
eventually leads to fracture.

\section*{Conclusions}

In this paper we have investigated the phenomenon of electron trapping in
the corundum (\alumina) lattice. Our simulation revealed that electrons
can only be trapped at temperatures lower than $T_d=250~K$. This is in good
agreement with experimental results obtained using the mirror method.
Trapping turns out to be possible only at lattice sites where the potential
surface of the uncharged lattice exhibits a minimum. This is in agreement
with the space charge model, as these sites appear to be polarizability
defects due to crystallographic dissymetry. Although our simulations
reproduce the main features of electron trapping at a semiquantitative
level, further improvement is needed in two directions: (a) modelling of
the short-range interaction of the electron with the ionic species, and
(b) quantum treatment of the electronic dynamics. Work is in progress along
these two lines.

The authors are grateful to E. Smargiassi for many useful suggestions, in
particular on the establishment of the shell model.

\section*{Table Captions}
\vspace{1 truecm}
\noindent\underbar{Table I}: Values of the parameters used for the
short-range interactions (* values from Catlow et al.~\cite{catlow}).

\vspace{0.5 truecm}

\begin{tabular}{|c|c|c|c|} \hline\hline
Interactions & A$_{\rm ij}$ (eV) & $\rho$ (\AA$^{-1}$) & C$_{\rm ij}$
(eV~\AA$^6$)  \\ \hline
Al-Al &   0*     &   0.1*     &  0*     \\
 O-O  &  22764*  &   0.149*   &  27.88* \\
Al-O  &  1460.3* &   0.29912* &  0*     \\ \hline\hline
\end{tabular}

\vspace{1 truecm}

\noindent\underbar{Table II}: Values of the parameters used for the
shell model (* values from Catlow et al..~\cite{catlow}).

\vspace{0.5 truecm}

\begin{tabular}{|c|c|c|} \hline\hline
Parameters & Al$^{3+}$ & O$^{2-}$ \\ \hline
spring constant (eV~\AA$^{-2})$ &   924.88    &  103.07*   \\
core charge (e)                 &  1.6170*    &  0.8106*   \\
shell charge (e)                &  1.3830*    & -2.8106*   \\ \hline\hline
\end{tabular}

\begin{figure}
\caption{Mean square displacement for the Al$^{3+}$ ions at T=300 $K$
and T=2170 $K$. The curves are scaled by a factor $x=0.01$ and $x=0.1$,
respectively.}
\end{figure}
\begin{figure}
\caption{Mean square displacement for the O$^{2-}$ ions at T=300 $K$
and T=2170 $K$. The curves are scaled by a factor $x=0.01$ and $x=0.1$,
respectively.}
\end{figure}
\begin{figure}
\caption{Potential surface for the uncharged lattice.}
\end{figure}
\begin{figure}
\caption{Path of the excess electron at $T=300~K$.}
\end{figure}
\begin{figure}
\caption{Path of the excess electron at $T=200~K$.}
\end{figure}
\begin{figure}
\caption{Mean square displacement of the electron at $T=300~K$ and
$T=200~K$.}
\end{figure}
\begin{figure}
\caption{Comparison between the potential surfaces of the uncharged
lattice and the charged one.}
\end{figure}
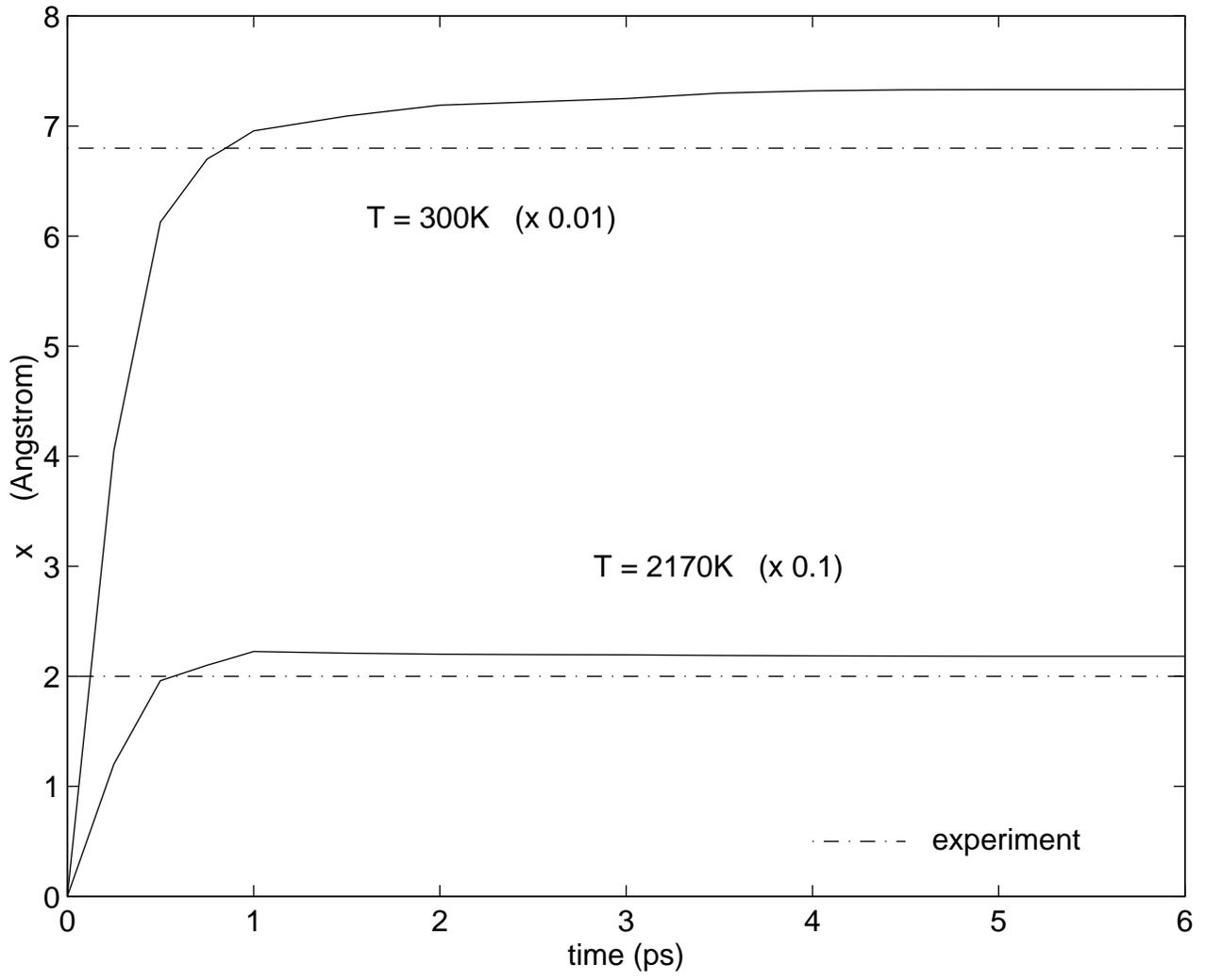
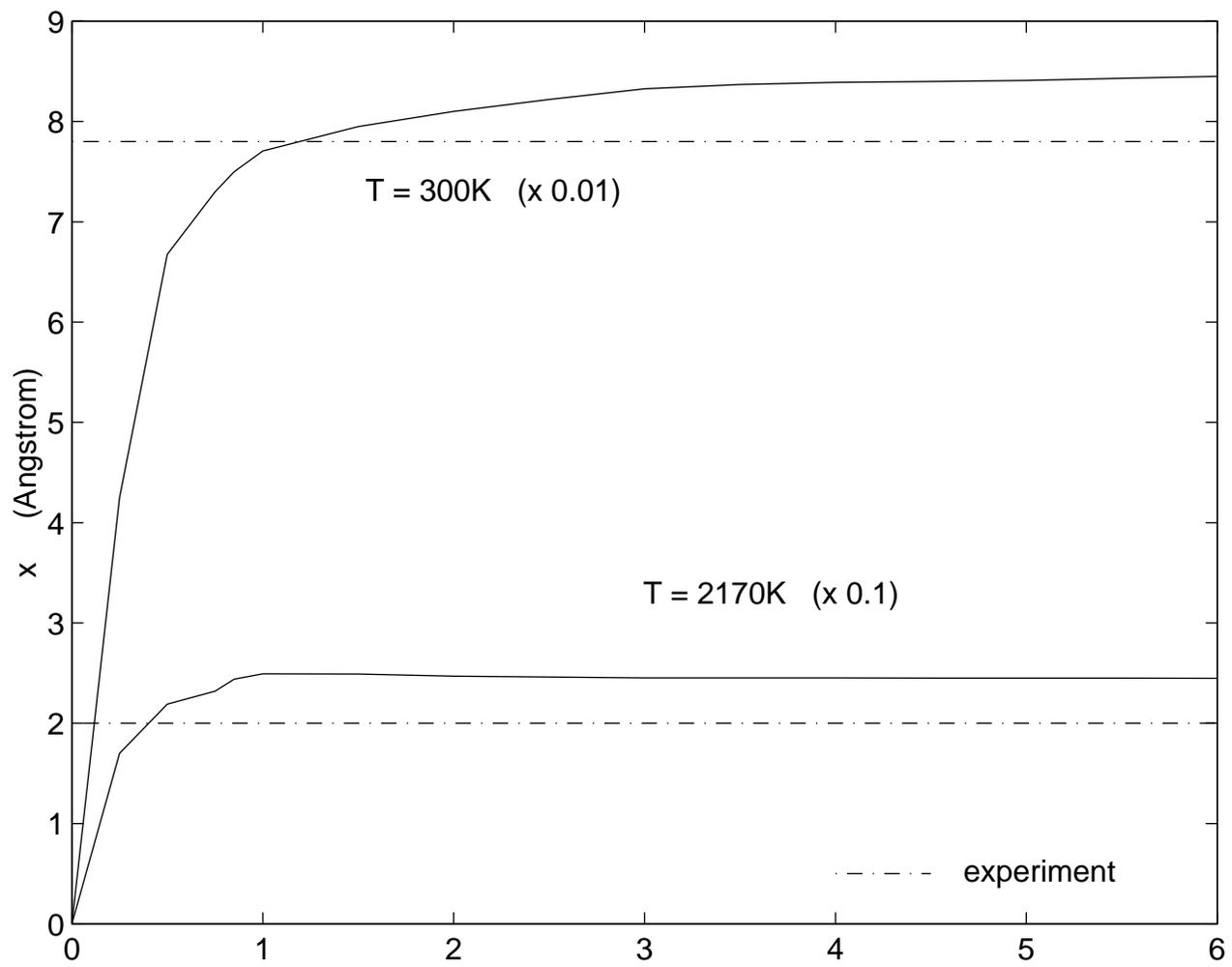
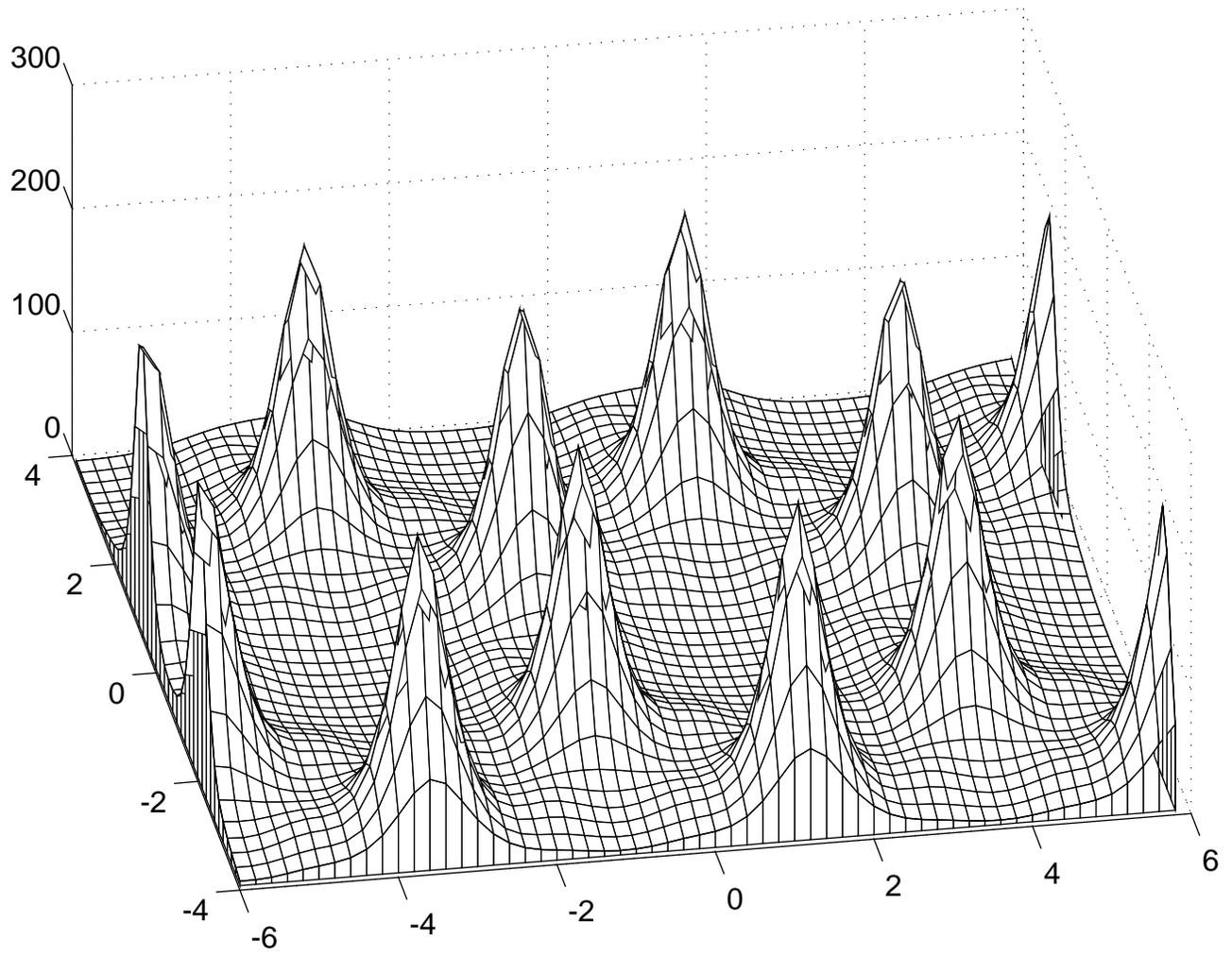
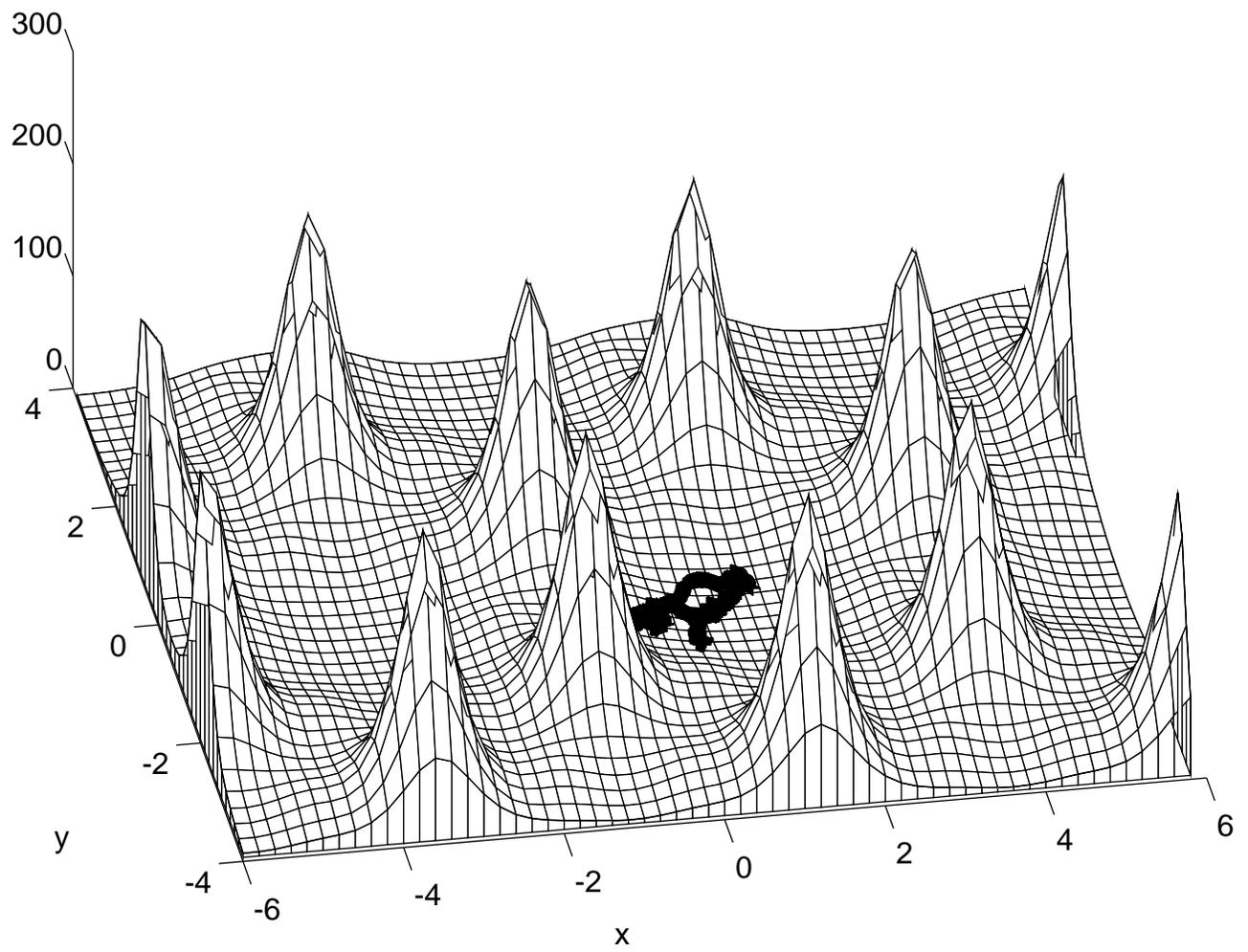
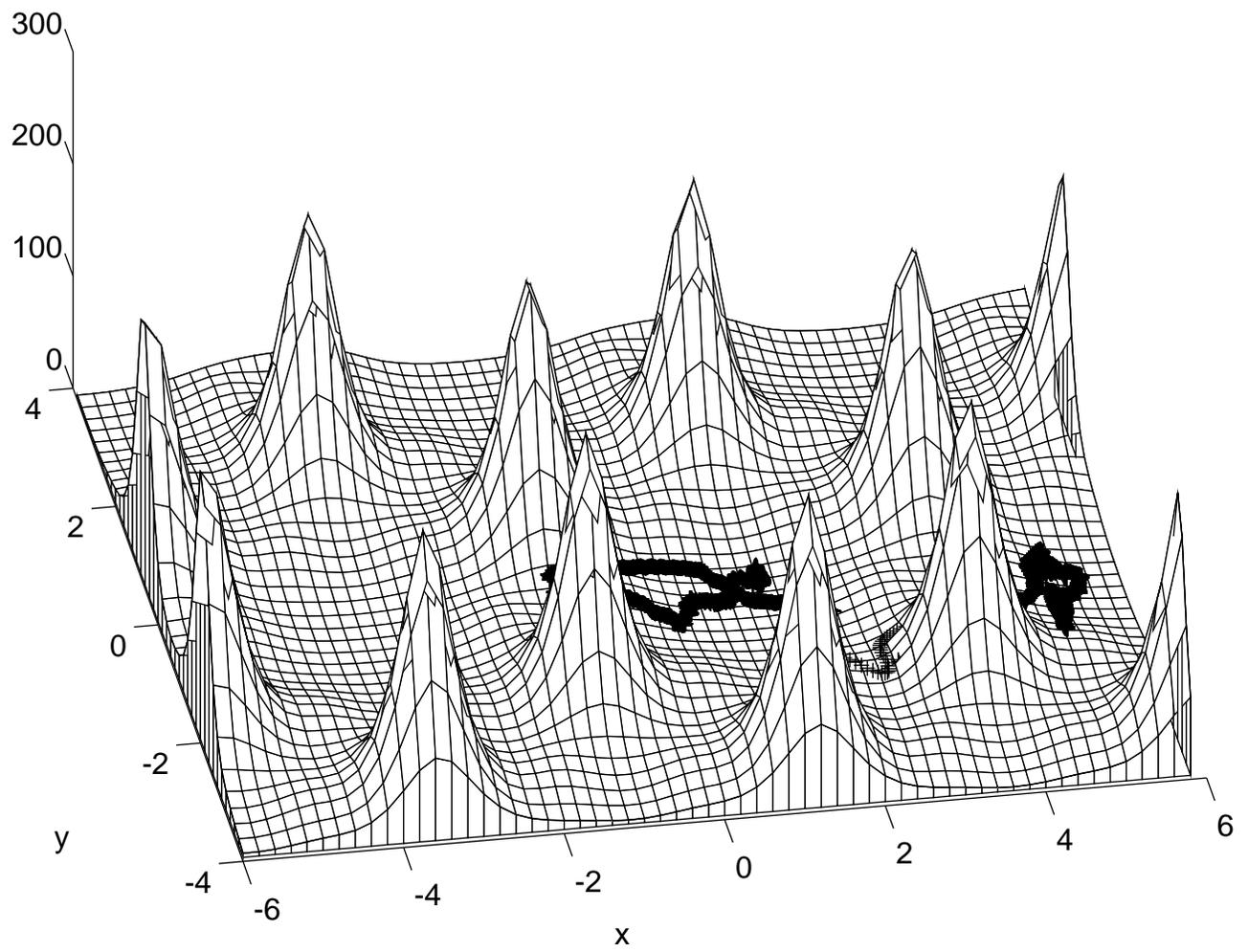
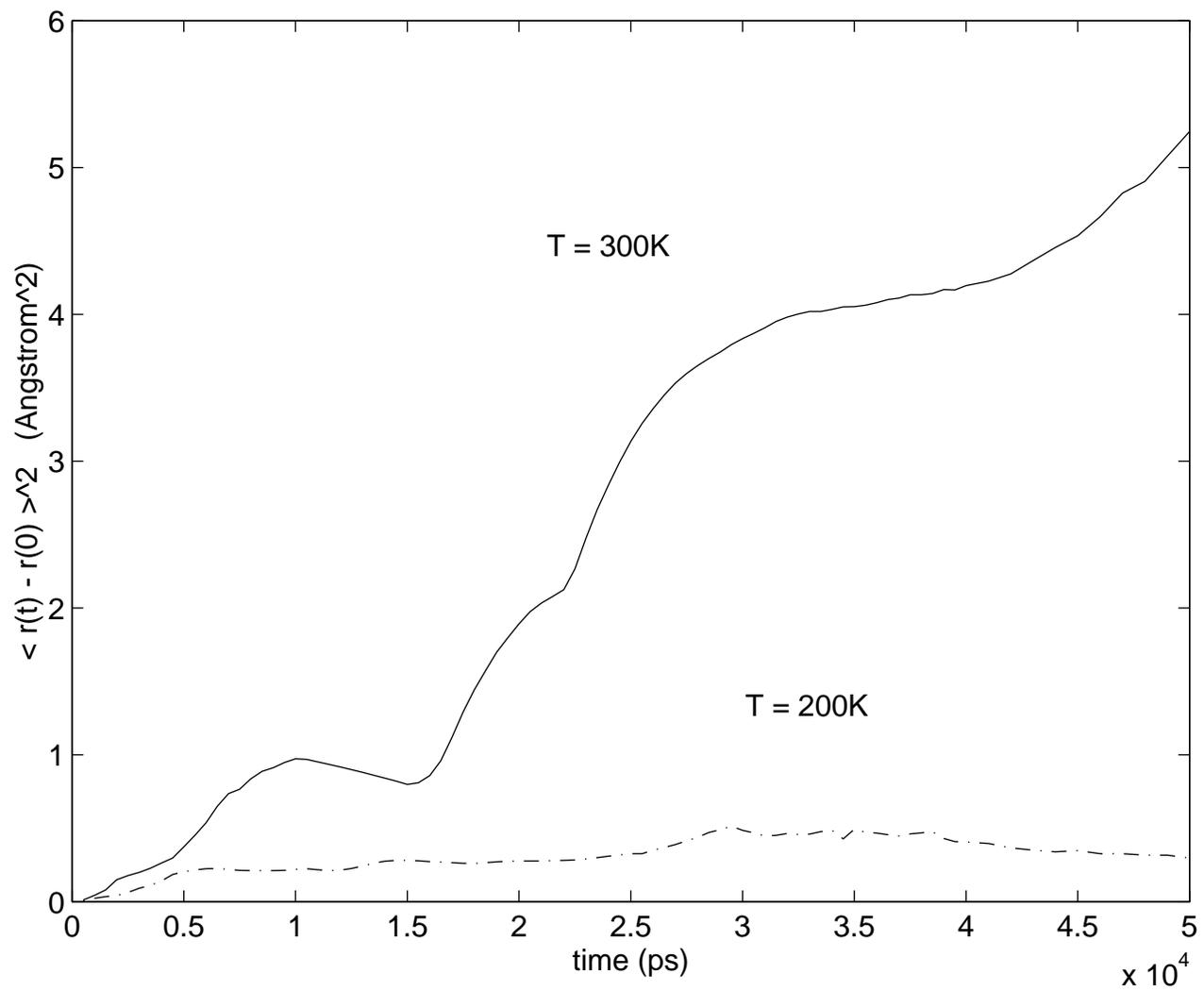
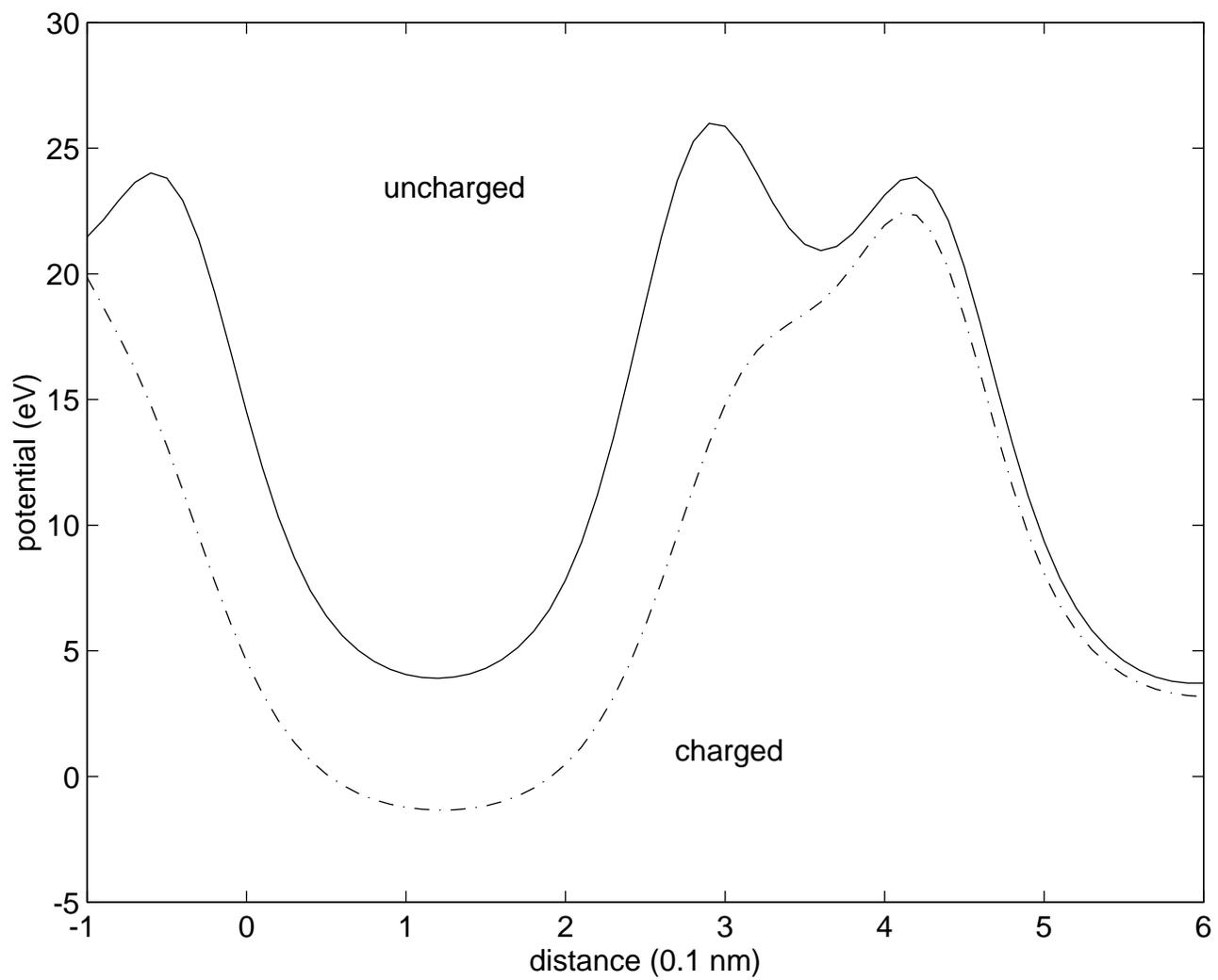
\end{document}